\documentstyle[12pt,aps]{revtex}

\newcommand {\ignore}[1]{}

\newcommand{\bc}{\begin{center}}
\newcommand{\ec}{\end{center}}

\def\ifmath#1{\relax\ifmmode #1\else $#1$\fi}

\def\3quarter{{\textstyle{3 \over 4}}}

\def\lf{\leaders\hbox to 1em{\hss.\hss}\hfill}
\def\e6{$E(6)$}
\def\10{$SO(10)$}
\def\21{$SU(2) \otimes U(1) $}

\def\422{$SU(4) \otimes SU(2) \otimes SU(2)$}
\def\321{$SU(3) \otimes SU(2) \otimes U(1)$}
\def\ne{\hbox{$\nu_e$ }}
\def\nm{\hbox{$\nu_\mu$ }}
\def\nt{\hbox{$\nu_\tau$ }}
\def\ns{\hbox{$\nu_{s}$ }}

\def\eq#1{{eq. (\ref{#1})}}

\def\lsim{\raise0.3ex\hbox{$\;<$\kern-0.75em\raise-1.1ex\hbox{$\sim\;$}}}
\def\gsim{\raise0.3ex\hbox{$\;>$\kern-0.75em\raise-1.1ex\hbox{$\sim\;$}}}

\def\beq{\begin{equation}}
\def\eeq{\end{equation}}
\def\bef{\begin{figure}}
\def\eef{\end{figure}}
\def\bet{\begin{table}}
\def\eet{\end{table}}
\def\bea{\begin{eqnarray}}
\def\ba{\begin{array}}
\def\ea{\end{array}}
\def\bi{\begin{itemize}}
\def\ei{\end{itemize}}
\def\ben{\begin{enumerate}}
\def\een{\end{enumerate}}

\def\ot{\otimes}
\def\eea{\end{eqnarray}}
\def\pr#1#2#3{           {\it Phys. Rev. }{\bf #1} (19#2) #3}

\def\n.c.#1#2#3{         {\it Nuovo Cim. }{\bf #1} (19#2) #3}
\def\r.n.c.#1#2#3{       {\it Riv. del Nuovo Cim. }{\bf #1} (19#2) #3}

\def\be{\begin{equation}}        
\def\ee{\end{equation}} 
\def\bear{\be\begin{array}}       
\def\eear{\end{array}\ee} 
\def\bea{\begin{eqnarray}} 
\def\eea{\end{eqnarray}}

\def\beqa{\begin{eqnarray}}
\def\eeqa{\end{eqnarray}}

\def\beq{\begin{equation}}
\def\eeq{\end{equation}}
\def\ba{\begin{array}}
\def\ea{\end{array}}

\def\21{$SU(2) \ot U(1)$} 
\def\ot{\otimes}

\begin{document}
\draft



  
  \title{Light Sterile Neutrino from extra dimensions and
    Four-Neutrino Solutions to  Neutrino Anomalies} \vskip 0.5cm
\author{A. Ioannisian
\footnote{On leave from Yerevan Physics Institute, Alikhanyan Br.2,
  Yerevan, 375036, Armenia.} and 
J. W. F. Valle} 
\address{Instituto de F\'{\i}sica Corpuscular -- C.S.I.C. \\
  Departamento de F\'{\i}sica Te\`orica -- Univ. de
  Val\`encia \\
  Edificio Institutos de Paterna -- Apartado de Correos 2085 --
  46071, Val\`encia, Spain
  http://neutrinos.uv.es }
\maketitle

\begin{abstract}
We propose a four-neutrino model which can reconcile the existing data
coming from underground experiments in terms of neutrino oscillations,
together with the hint from the LSND experiment and a possible
neutrino contribution to the hot dark matter of the Universe.  It
applies the idea that extra compact dimensions, probed only by gravity
and possibly gauge-singlet fields, can lower the fundamental scales
such as the Planck, string or unification scales.  Our fourth light
neutrino $\nu_s$ ($s$ for sterile) is identified with the zero mode of
the Kaluza-Klein states. To first approximation \ns combines with the
\nm in order to form a Dirac neutrino with mass in the eV range
leaving the other two neutrinos massless.
The smallness of this mass scale (suitable for LSND and Hot Dark
Matter) arises without appealing neither to a see-saw mechanism nor to
a radiative mechanism, but from the volume factor associated with the
canonical normalization of the wave-function of the bulk field in the
compactified dimensions.
On the other hand the splitting between \nm and \ns (atmospheric
scale) as well as the mass of the two other neutrinos (solar mass
scale) arise from the violation of the fermion number on distant
branes.
We also discuss alternative scenarios involving flavour-changing
interactions. In one of them \ne can be in the electron-volt range and
therefore be probed in beta decay studies.
\end{abstract}
\newpage

\section{Introduction}

It was recently suggested \cite{Antoniadis:1990ew,ADD1,AADD,ADD11}
that the hierarchy problem (the smallness of the weak scale to the
Planck scale) may be avoided by simply removing the large scale. The
observed small value of gravitational constant at long distance is
ascribed to the spreading of the gravitational force in $n$ extra
spatial dimensions.  The relation between the scales where gravity
becomes strong in the 4+n dimensional theory can be derived from Gauss
law
\beq
\label{gauss}
M^2_{Pl}\simeq(R \ M_F)^n M_F^2 ,
\eeq
where $R$ is the compactification radius of the additional dimensions
and $M_F$ is the fundamental Planck scale, which can be low. In what
follows we take $M_F \simeq 10$ TeV and $n=6$ for which the
corresponding value of $R$ is $R \simeq 10^{-12}$ cm.  It follows that
the Standard Model (SM) fields (quarks, leptons, gauge fields and
possible Higgs multiplets) are confined to a brane configuration
\cite{L1}, while the large compactified dimensions are probed only by
gravity and {\sl bulk} fields \cite{ADD1}, singlet under the \321
gauge group.  It has been recently shown \cite{L1} that this framework
can be embedded into string models, where the fundamental Planck scale
can be identified with the string scale which could be as low as the
weak scale. The extra dimensions have the potential to lower the
unification scale as well \cite{DDG1}.

In Refs.\cite{DDG,ADDM} it was shown how neutrinos can naturally get
very small {\sl Dirac} masses via mixing with a bulk fermion. The main
idea is that the coupling of the bulk fermions to the ordinary
neutrino $\nu$ is automatically suppressed by a volume factor
corresponding to the extra compactified dimension. Such a volume
factor arises from the canonical normalization of the wave-function of
the bulk field in the compactified dimensions.  This volume factor
provides a natural mechanism for suppressing the Yukawa coupling and
correspondingly yields a light neutrino mass.  In these theories, the
left-handed neutrinos as well as other standard model (SM) particles,
are localized on a brane embedded in the bulk of the large extra
space.
In Refs.\cite{ADDM,AD} it was shown how to generate small {\sl
  Majorana} masses for neutrinos via strong breaking fermion number on
distant branes.

In this paper we study the implications of theories with extra
dimensions with a few TeV scale of quantum gravity for neutrino
physics. In contrast with other attempts to study neutrino masses and
oscillations in models with large extra dimensions ~\cite{smiv99} we
consider the possibility of using these ideas in order to find
possible realistic ways to account for all the present neutrino
observations from underground experiments\footnote{ For a recent
  updated global analysis of solar and atmospheric neutrino data
  including the solar neutrino recoil electron day and night spectra
  (for possible seasonal effects see~\protect\cite{deHolanda:1999ty}),
  as well as the atmospheric up--going muon data see
  ref.~\protect\cite{Gonzalez-Garcia:2000sq}. For previous refs. see
  ref.~\protect\cite{Gonzalez-Garcia:1999aj} and
  ref.~\protect\cite{atm00}.}, including both solar and atmospheric
data, in terms of neutrino oscillations, in addition to the
accelerator data, namely the possible hint from the LSND experiment.
The latter requires at least one light sterile neutrino, in addition
to the three (active) neutrinos. We propose four-neutrino models which
explain the smallness of neutrino masses by exploiting the above
mechanisms.  In our model the sterile neutrino survives from the
neutral fermions of the bulk sector, being identified with the zero
mode of the Kaluza-Klein states. Thus we bypass the need for a
protecting symmetry to justify its lightness. To first approximation
the sterile neutrinos and one of the active ones combine to form a
Dirac state, while the other two remain massless.  Next, as a result
of fermion number breaking on distant branes, the heavier states split
into a Quasi--Dirac state ~\cite{Valle:1983dk} at the eV scale, while
the others get a small mass.  
Our simplest scheme obtained this way reproduces the model proposed in
ref.~\cite{PTV92}, with the atmospheric neutrino data accounted for
due to maximal mixing \nm to \ns oscillations, while the solar
neutrino data is explained through \ne to \nt MSW conversions.

Since, although allowed \cite{atm00}, the \nm to \ns oscillation
channel is not the preferred explanation of the atmospheric neutrino
anomaly, we also briefly discuss the role of flavour-changing (FC)
transitions. In this context we also mention, for example, a new
variant of the model given in ref.~\cite{PV93} which is, however,
physically inequivalent. In this new model the Quasi--Dirac neutrino
with mass at the LSND scale combines \ne with the sterile neutrino
\ns, leaving \nm and \nt massless to first approximation. As a result
the model leads to possible effects in tritium beta decay and the
atmospheric neutrino data are well explained due to \nm to \nt
conversions.  In contrast to the first, this second scenario may
successfully explain the solar neutrino data, but not via the MSW
mechanism: it requires the presence of other physical processes such
as flavour-changing (FC) transitions. These have been considered both
in the context of solar neutrinos, as well of atmospheric neutrinos
~\cite{fc}.

\section{The Simplest Model}

Here we concentrate in the minimum brane-inspired scheme in which all
elements required to explain the neutrino anomalies (the LSND/HDM as
well as the solar and atmospheric mass scales) are generated by the
physics of extra dimensions. For definiteness, we will be considering
a model that minimally extends the standard field content by one bulk
neutrino, $N(x,y)$, singlet under the \321 gauge group. This
propagates on a $[1+(3+\delta)]$-dimensional Minkowski space with
$\delta \le n$.  Each $y$-coordinate of the large dimensions is
compactified on a circle of radius $R$, by applying the periodic
identification: $y \equiv y + 2\pi R$.  Furthermore, we consider that
only one four--dimension \321 singlet $\xi(x,y)$ from the
higher-dimensional spinor $N(x,y)$ has non-vanishing Yukawa couplings,
$\bar{h}_l$, to the three ordinary isodoublet leptons $L_l (x)$, with
$l = e, \mu, \tau$.  The relevant Lagrangian has the form

\begin{eqnarray}
\label{Leff}   
{\cal L}_{\rm eff}  & =&  \int \!\! dy^{\delta}\
 \Big[\, \bar{N} \Big( i\gamma^\mu \partial_\mu\, +\,
 i\gamma_{\vec{y}} \partial_{\vec{y}} \Big) N\ +\ \delta (y)\,
\Big( \sum_{l=e,\mu,\tau}\, \bar{h}_l L_l\tilde{\Phi} \xi\, +\, {\rm
H.c.}\,\Big) +\, \delta (y)\, {\cal L}_{\rm SM}\,\Big]\, ,
\end{eqnarray}
where  $\tilde{\Phi} = i\sigma_2 \Phi^*$ and  ${\cal L}_{\rm SM} $
describes the SM  Lagrangian.   The dimensionful Yukawa couplings
$\bar{h}_l$ may be related to the dimensionless ones, $h_l^{(\delta)}$, 
through
\begin{equation}
  \label{hl}
\bar{h}_l\ =\ \frac{h_l^{(\delta)}}{(M_F)^{\delta /2}}\ .
\end{equation}

We can now express the 4+$\delta$-dimensional two-component spinor
$\xi$ in terms of a Fourier series expansion as follows:
\begin{eqnarray}
  \label{xi}
\xi (x,y) &=& \frac{1}{(2\pi R)^{\frac{\delta}{2} }}\ \sum_{\vec{n}}
\xi_{\vec{n}}(x)\
                   \exp\bigg(\frac{i\vec{n}~.~\vec{y}}{R}\bigg)\,
\end{eqnarray}
Substituting Eq.\ (\ref{xi}) into the effective Lagrangian
(\ref{Leff}) then performing the $y$ integrations and integrating out
heavy Kaluza-Klein states yields
\be
  \label{LKK}   
{\cal L}_{\rm eff}  =  {\cal L}_{\rm SM} \ 
+\, \Big(\, \sum_{l=e,\mu,\tau}\,h_l\, L_l\tilde{\Phi} \xi_0\  +\
{\rm H.c.}\,\Big)\, ,
\end{equation}
where $\xi_0$ is the zero mode of the Kaluza-Klein states and
\begin{equation} 
  \label{hln}
h_l\ =\  \bigg(\frac{M_F}{M_{\rm P}}\bigg)^{\frac{\delta}{n}}\
h_l^{(\delta)} .
\end{equation}

As was first noticed in \cite{DDG,ADDM}, the four-dimensional Yukawa
couplings $h_l$ are naturally suppressed by the volume factor
$\bigg(\frac{M_F}{M_{\rm P}}\bigg)^{\frac{\delta}{n}}$ of the extra
dimensions.  With this we can estimate the effective four-dimensional
Yukawa coupling and the corresponding neutrino mass in our model. In
order to account for the LSND or hot dark matter mass scale $m_\nu
\sim 1$ eV or so, we choose $\delta=4$ and $n=6$, giving $h_l \sim
10^{-10} h_l^{(\delta)}$.

Now we turn to the Majorana masses for neutrinos. These are crucial in
order to generate the mass splittings required in neutrino oscillation
interpretations of the solar and atmospheric neutrino anomalies found
in underground experiments. As shown in Refs.~\cite{ADDM,AD} the
neutrinos may get small Majorana masses via interactions with distant
branes where fermion number is maximally broken.  In case it is
assumed that these interactions proceed via a very light field
(lighter than $\sim 1/R$ but heavier than $\sim ($mm$)^{-1}$ to have
escaped detection) which propagates in 4+$\delta$ dimensions and that
the brane where lepton number is broken is as far away as possible
i.~e.  at a distance $\sim R$.  The Majorana part of the neutrino
masses is then expected to be 
\be
\label{majorana}
m_{l l^\prime} \sim f_{l l^\prime}\frac{v^2}{M_F}
\left(\frac{M_F}{M_{\rm
pl}}\right)^{2\delta/n - 4/n}
\label{light}
\ee

The neutrino mass matrix takes in the basis ($\nu_e, \nu_\mu, \nu_\tau,
\nu_s$ ) ($\nu_s=\xi_0$) the form

\beq
      \cal M_{\nu} ~=~ \pmatrix{
         m_{l l^\prime} &  M_l           \cr
         M^T_{l^\prime}     &  0     \cr}~.
\label{massmatrix}
\eeq Here $M_l = h_l v$, where $v$=174 GeV is the standard electroweak
vacuum expectation value. Note that the $m_s$ entry in \eq{massmatrix}
has been omitted 
\footnote{Note that one would have to add a term
    \begin{displaymath}
           \pmatrix{
         0 &  M^\prime           \cr
         M^\prime     &  M_F     \cr}~. \nonumber
  \end{displaymath}
in the basis ($\nu_s$, $\chi$), where $\chi$ denotes a fermion on a
distant brane which couples with our bulk field \ns.  Here $M_F \sim $
TeV is the effective Planck mass and $ M^\prime \sim$ eV is suppressed
by the same volume factor as $M_l$. Thus this would lead to an
effective $m_s$ entry in \eq{massmatrix} of order $10^{-12}$ eV.
Clearly this is totally irrelevant for our problem.}
since the bulk sector where the sterile neutrino \ns
lives is eight--dimensional (see Sec.2).  There is, however, a general
theorem which states that in eight dimensions there can be no massive
Majorana spinor~\cite{thm}.

In the limit that Dirac mass terms ($M_l$) are much bigger then
Majorana mass terms ($m_{l l^\prime}$) two of the neutrinos are
massless and other two form Dirac state with a mass

\beq
\label{dirac}
m \equiv m_{LSND/HDM} = \sqrt{M^2_e+M^2_\mu+M^2_\tau}  
\eeq

This state is identified by two angles $\theta$ and $\varphi$ defined as

\beq
\label{thetaphi}
\sin \theta = \frac{M_e}{m} , \ \ \tan \varphi = \frac{M_\mu}{M_\tau}. 
\eeq

The entries $m_{l l^\prime}$ only arise due to the breaking lepton
number on distant branes. In the case $\delta = 4$ and $n=6$ they are
suppressed compared to the Dirac mass terms by the factor
$\frac{v}{M_F}\frac{f_{l l^\prime}}{h_l }$.  These terms give masses
to the lowest-lying neutrinos and also responsible for splitting Dirac
state to two Majorana states. For suitable values of the parameters,
these are in the right range to have a solution for solar and
atmospheric neutrino deficit. More especifically, from the latest fits
one needs \cite{atm00}
\beq
\Delta m^2_{atm} \simeq 3.5 \times 10^{-3} eV^2
\eeq
in order to account for the full set of atmospheric neutrino data.

On the other hand the latest global analysis of solar neutrino data
slightly prefers the large mixing MSW solution (LMA) characterized by
the best-fit point \cite{Gonzalez-Garcia:1999aj}
\beq
\Delta m^2_{LMA} \simeq 3.6 \times 10^{-5} eV^2
\eeq
Now assuming naturalness, namely, that masses and splittings are of
the same order $\Delta m_{atm} \simeq m_{l l^\prime} \simeq
\sqrt{\Delta m^2_{\odot}}$ and since $\Delta m^2_{atm} \simeq 2
m \Delta m_{atm}$, one finds
\beq
m \simeq 0.8 eV
\eeq
characterizing the order of magnitude of the LSND/HDM scale in the LMA
case.

On the other hand, for the SMA solution we have
\beq
\Delta m^2_{SMA} \simeq 5 \times 10^{-6} eV^2
\eeq
so that if $\Delta m^2_{SMA} \simeq (\Delta m)^2$ and using again
$\Delta m^2_{atm} \simeq 2 m \Delta m_{atm} $  one finds
\beq
m \simeq 0.3 eV
\eeq
characterizing the order of magnitude of the LSND/HDM scale in the SMA
case.

Of course since clearly the solar mass splitting need not coincide
exactly with the lightest state masses, the above estimates are meant
to be crude order-of-magnitude estimates only. As a result the
LSND/HDM scales both in the LMA and in the SMA case can be larger than
estimated above.

In the above approximation the form of the charged current weak
interaction may be given as 
\be
\label{}
-\frac{g}{\sqrt{2}}\sum_{i=1}^3 \sum_{\alpha=1}^4
\bar{e}_{iL}\gamma^\mu K_{i\alpha}\nu_{\alpha L} + h.c.
\ee
where
\beq
      \cal K ~=~ \pmatrix{
c_\theta c_m &  c_\theta s_m   &  \frac{s_\theta}{\sqrt{2}} & \frac{s_\theta}{\sqrt{2}} \cr
- s_m c_\varphi - s_\theta s_\varphi c_m  &  c_\varphi c_m-s_\theta s_\varphi s_m 
               & \frac{c_\theta s_\varphi}{\sqrt{2}} 
                     &  \frac{c_\theta s_\varphi}{\sqrt{2}} \cr
  s_m s_\varphi-s_\theta c_\varphi c_m & -s_\varphi c_m-s_\theta c_\varphi s_m 
               &  \frac{c_\theta c_\varphi}{\sqrt{2}} &  \frac{c_\theta c_\varphi}{\sqrt{2}} \cr
  0    &  0    & -\frac{1}{\sqrt{2}}     &  \frac{1}{\sqrt{2}}                  \cr}~.
\label{mixingmatrix}
\eeq 
Here the first, second and third rows denote $\nu_e$, $\nu_\mu$ and
$\nu_\tau$ respectively, while the fourth is the sterile neutrino
$\nu_s$. The angles $\theta$ and $\varphi$ identify the dark matter
neutrino while $\theta_m$ diagonalizes the light sector. The matrix
$K$ determines also the structure of the neutral current weak
interactions~\cite{Schechter:1980gr}
\beq
- \frac{g'}{2 \sin \theta_W}
Z_{\mu} \:
\sum_{\alpha \beta} \bar{\nu}_{L\alpha} \ \gamma_{\mu} \ P_{\alpha\beta} \
\nu_{L\beta}
\label{NC2}
\eeq
through the relation
\beq
P = {\cal K}^\dagger {\cal K}
\label{NC1}
\eeq

\section{Phenomenology}

If we assume that muonic neutrino coupling to the high dimensional
spinor is dominant ($h_e, h_\tau \ll h_\mu \simeq 0.1$ ) the light
sterile neutrino, $\nu_s$, combines with $\nu_\mu$ and form a
Quasi--Dirac state, crucial to account for the hint coming from the
LSND experiment, and may also contribute to the hot dark matter of the
Universe.
Apart from the mass of the Quasi--Dirac state~\cite{Valle:1983dk}, one
has the splittings between its components, as well as the masses of
two light active states.
The splittings between the heavy states and that characterizing the
lighter neutrinos will be associated with the explanations of the
atmospheric and solar neutrino anomalies, respectively.
The atmospheric neutrino deficit is ascribed to the $\nu_\mu$ to
$\nu_s$ oscilations.  The solar neutrino problem could be solved via
MSW small or large angle $\nu_e$ to $\nu_\tau$ solutions. This
reproduces exactly the phenomenological features of the model proposed
in ref.~\cite{PTV92}, providing a complete scenario for the present
neutrino anomalies. 

From eq.(\ref{mixingmatrix}) we can determine the pattern of neutrino
oscillations predicted in the model.  \textit{In vacuo} the neutrino
oscillation probabilities are simply given as
due to the cumulative non-decoupling effects of the Kaluza-Klein
neutrinos in electroweak processes

\begin{eqnarray}
  \label{eq:oscil}
 P(\ne \to \nm) &=& \sin^2 2\theta  \sin^2 \varphi ~
~\sin^2 \frac{m^2 L}{4E} \\    \nonumber
 P(\ne \to \nt) &=& \sin^2 2\theta  \cos^2 \varphi ~
~\sin^2 \frac{m^2 L}{4E} \\  \nonumber
 P(\nm \to \nt) &=& \sin^2 2\varphi \cos^4 \theta  ~
~\sin^2 \frac{m^2 L}{4E}  \nonumber
\end{eqnarray}
One sees that the first consequence of our model is to have
potentially detectable rates for neutrino oscillations in the
laboratory. The LSND experiment gives the following constraints
$\sin^2 2\theta \sin^2\varphi \simeq 0.003$ (0.03) for $m \simeq 1.0
(0.5)$ eV.  
On the other hand the Chooz experiment requires $\sin^2 2\theta \le
0.18$ for $m \sim 0.5 \: - \: 1$ eV.

Note that the eV range sterile neutrino range with maximal mixing with
the active state predicted in our scenario would enter into
equilibrium with the active neutrinos via neutrino oscillations in the
early Universe~\cite{bbnsterile}, leading to an effective equivalent
number of four light neutrinos in the early Universe, $N^{eff}_\nu=4$.
For recent discussions see ref.~\cite{sarkar,Fiorentini:1998fv} and
for a discussion of the possible role of lepton asymmetries see ref.
~\cite{Foot:1997qc}.  This contrasts with the situation found in the
model of ref.~\cite{PV93}, since there the lightness of the sterile
neutrino suppresses these conversions leading to an effective light
neutrino number of three, $N^{eff}_\nu=3$.  The difference lies in
that in the present case the Quasi--Dirac state (of mass $m \sim$ eV,
for $M_F \simeq 10$ TeV) inescapably includes the sterile neutrino,
while in ~\cite{PV93} the Quasi--Dirac state combines the two active
\nm and \nt \cite{Wolfenstein:1981kw} leaving the \ns in the lighter
sector.

Finally, note that the low value for the scale $M_F \simeq $ few TeV
might lead to flavour-violation and universality-breaking phenomena
potentially accessible to experiment due to the cumulative
non-decoupling effects of the Kaluza-Klein neutrinos in electroweak
processes, even though the original Yukawa couplings are
small~\cite{Ioannisian:1999cw}.
The rates of the flavour-violating decays $\mu \to e \gamma$ and $\mu
\to eee$ as well as $\mu \to e$ coherent conversion in nuclei are
proportional to the Yukawa couplings 
\be h_e^2 h_{\mu}^2 = \frac{m^4}{4v^4} \sin^2 2 \theta \sin^2\varphi
\ee
On the other hand the $\nm \to \ne $ oscillation probability
\eq{eq:oscil} depends on the same parameters.  Fitting for the LSND
experiment and using the upper limits from non-observation of lepton
flavour-violating and universality-breaking phenomena involving the W
and Z bosons one obtains \cite{Ioannisian:1999cw} the following
restriction on the compactification scale $M_F$ in our model
\be
M_F \gsim 2  \textrm{TeV} \ \left(\frac{m}{\textrm{eV}}\right)^{3/5}
\left(\frac{\sin^2 2 \theta \sin^2\varphi}{0.003} \right)^{3/20} 
\ee
Note that in the region indicated by the LSND experiment the product
of the two terms in parenthesis in the right hand side are essentially
constant and close to unity. With this one can obtain a lower bound on
the scale $M_F$, $M_F \gsim $ 2 TeV. Note that this is an order of
magnitude estimate only, since we do not know where to cut the
Kaluza-Klein tower of the sterile neutrinos.

There are other conceivable brane-inspired four--neutrino scenarios
capable of explaining of the solar and atmospheric neutrino problems.
For example, if we assume that the electronic neutrino coupling to the
high dimensional spinor is the dominant one ($h_\mu, h_\tau \ll h_e
\simeq 0.1$ ) then the Quasi--Dirac state combines $\nu_e$ with
$\nu_s$.  This leads to the possibility of observing neutrino mass
effects in tritium beta decay experiments and also the \ne would form
part of the hot dark matter.  The explanation of atmospheric neutrino
deficit comes due to the $\nu_\mu$ to $\nu_\tau$ oscillations, which
gives an excellent fit of the data ~\cite{atm00}, better than that for
the \nm to \ns case.  In contrast, the solar neutrino problem can not
be accounted for in the framework of the MSW effect, since in this
case we have large mixing angle $\nu_e$ to $\nu_s$ oscillations, which
is ruled out by the solar data fit~\cite{Gonzalez-Garcia:1999aj}. A
possible way out is to assume the presence of other physical
processes, such as flavour-changing (FC) transitions.  In this case
one may have fermion number violation on distant branes in such a way
that only the \nm and \nt masses are split, but not the heavier
neutrinos \ne and \ns.  This can provide an explanation of the solar
neutrino data in the presence of flavour-changing neutral
currents~\cite{fc}.

\section{Discussion}

We have proposed a four-neutrino model which can explain the smallness
of neutrino masses without appealing neither to a see-saw mechanism
nor to a radiative mechanism. It applies the idea that extra compact
dimensions, probed only by gravity and gauge-singlet fields, can lower
the fundamental scales such as the Planck, string or unification
scales.  The light sterile neutrino $\nu_s$ is identified with the
zero mode of the Kaluza-Klein states and combines with \nm in order to
form a Quasi--Dirac neutrino with mass in the eV range, contributing
to the hot dark matter of the Universe.  Apart from the possible
effect of lepton asymmetries the sterile neutrinos are brought into
equilibrium in the early Universe, leading to four effective light
neutrinos $N^{eff}_\nu=4$.  Fermion number breaking effects on distant
branes lead to the splitting of this Quasi--Dirac state as well as two
massless active states. These splittings generate oscillations which
may reconcile the existing data coming from underground experiments
with the hint from LSND.
The simplest scheme is the same as the one proposed in
ref.~\cite{PTV92}, accounting for the atmospheric data in terms of \nm
to \ns oscillations and the solar neutrino data in terms of \ne to \nt
MSW conversions.

The scheme can be tested through lepton flavour violating processes
such as $\mu \to e + \gamma$, $\mu \to 3 e$ and muon conversion in
nuclei. In particular we showed that fitting for the LSND signal
enables us to estimate the muon-number violating rates as a function
of the fundamental Planck scale, $M_F$ which must exceed few TeV or
so.
Finally, note that in our scheme there is no expected modification of
the Newtonian law accessible at the proposed sub-millimeter scale.

We have also discussed a second possible model which accounts for the
atmospheric data in terms of \nm to \nt oscillations.  However, in
contrast to the original model of ref.~\cite{PV93}, such model is
incompatible with an explanation of the solar data in the framework of
the simplest MSW effect since the large mixing predicted between \ne
and \ns is ruled out by the solar neutrino
data~\cite{Gonzalez-Garcia:1999aj} \footnote{This follows from the
  behaviour of the neutrino conversion probability expected for large
  mixing combined with the absence of neutral currents in
  active-to-sterile conversions.}.
This can be remedied in the presence of flavour-changing (FC)
interactions.

Finally we comment on another recent attempt~\cite{Mohapatra:1999af}
to obtain a four-neutrino model based on extra dimensions.  The model
in ref.~\cite{Mohapatra:1999af} is totally different.  Ours is minimal
in the sense that we have simply the standard \321 gauge group and no
right--handed neutrinos, while they explain the lightness of the
active neutrinos via the standard left-right symmetric see-saw
mechanism. Moreover we have a much lower value for the fundamental
Planck scale in our model, $M_F \sim 10^4$ GeV. This leads to a
plethora of lepton flavour violating processes which are expected to
be negligible in the model of ref.~\cite{Mohapatra:1999af} and which
could potentially be used to distinguish them.  Finally, in our case
\emph{all} light neutrino masses arise from the extra dimensions
mechanisms: the eV scale follows from the volume factor associated
with the canonical normalization of the wave-function of the bulk
field in the compactified dimensions, while solar and atmospheric mass
scales arise from the violation of the fermion number on distant
branes.

\section{Acknowledgement}

We thank Subir Sarkar and Sasha Dolgov for discussions on the present
status of primordial nucleosynthesis bounds and their uncertainties.
This work was supported by DGICYT grants PB98-0693 and SAB1998-0136,
as well as the Alexander von Humboldt Foundation and the European
Union, under TMR contract ERBFMRX-CT96-0090.


\end{document}